\documentclass[%
 reprint,
nofootinbib,
 amsmath,amssymb,
 aps,
]{revtex4-1}

\usepackage{graphicx}
\usepackage{dcolumn}
\usepackage{bm}

\usepackage{amsmath}
\usepackage{amssymb}
\usepackage{amstext}

\begin{document}


\title{Non-local gravity in D-dimensions: Propagator, entropy and bouncing Cosmology}


\author{Aindri\'u Conroy~$^1$}
\author{Anupam Mazumdar~$^{1,~2}$}
\author{Spyridon Talaganis~$^{1}$}
\author{Ali Teimouri~$^{1}$}
\affiliation{
$^{1}$~Consortium for Fundamental Physics, Lancaster University, Lancaster, LA1 4YB, UK\\
$^{2}$~IPPP, Durham University, Durham, DH1~3LE, UK}


\begin{abstract}
We present the graviton propagator for an infinite derivative, $D$-dimensional, non-local action, up to quadratic order in curvature around a Minkowski 
background, and discuss the conditions required for this class of gravity theory to be ghost-free. We then study the gravitational entropy for de-Sitter and 
Anti-de Sitter backgrounds, before comparing with a recently derived result for a Schwarzschild blackhole, generalised to arbitrary 
$D$-dimensions, whereby the entropy is given simply by the area law. A novel approach of decomposing the entropy into its 
$(r,t)$ and spherical components is adopted in order to illustrate the differences more clearly. We conclude with a discussion of de-Sitter entropy in the framework of a non-singular bouncing cosmology.
\end{abstract}

\maketitle


\section{\label{sec:intro}Introduction}

Einstein's general relativity (GR) is a well-tested theory of gravity in the infrared
(IR)~\cite{Will}, reducing to Newtonian gravity for non-relativistic 
systems. At very short distances, the Newtonian potential has
been tested only up to the distance scale of  $10^{-5}$~m,  which is the IR limit
of Einstein's gravity for a slowly time varying test mass~\cite{Adelberger}. The corresponding
mass scale is around  $M\sim 1$~eV, which means that beyond this mass scale,
the nature of gravity itself is very poorly constrained.  As it stands, 
Einstein's theory is plagued by spacetime singularities -  be it, the singularity of a blackhole or the initial, cosmological singularity~\cite{Hawking:1973uf}. The latter is believed to be
a naked  singularity, which prevents, not only geodesic completeness for any null or
time like trajectories, but also prevents us from comprehending the initial classical and quantum state of the Universe.

Einstein's gravity, however, is unique in a sense that it predicts
the area-law of gravitational entropy for a gravitationally bound system.
In $4$-dimensions, the Bekenstein-Hawking entropy of a blackhole is given by the
area of a $2$-dimensional horizon~\cite{Bekenstein,Hawking}, which may also be confirmed by following
Wald's prescription for the gravitational entropy of a static, homogeneous and
isotropic metric, supplemented with a horizon~\cite{Wald}, see also \cite{Kang}. This observation has lead to many well-known
conjectures, such as the famous anti-deSitter
(AdS) conformal field theory (CFT) correspondence~\cite{Maldacena}, and
the \emph{holographic principle}, which states that the gravitational entropy
is proportional to the the surface area rather than the volume~\cite{thooft,Susskind}. The illusory nature of the volume and the subsequent holographic quality of the Universe, itself, affords the principle its name.   

Of course, some of the concerns raised by Einstein's gravity at short distances and small time scales need to be addressed  by better understanding the ultraviolet
(UV) aspects of gravity, while retaining the aforementioned predictions in the IR.  However,
the challenges in modifying gravity arises from two fronts. On one hand, higher-order curvature corrections tend to ameliorate the UV 
aspects by suppressing the graviton propagator. On the other, gravity being a derivative theory, 
higher-order corrections naturally introduce extra propagating states, which can be massive, tachyonic
or ghost-like. A classic example is Stelle's $4$th derivative gravity which is renormalisable, but contains a massive
spin-2 ghost~\cite{Stelle}.

In Ref.~\cite{Biswas:2011ar,Biswas:2013kla}, the issue of ghosts has been addressed at tree-level in $4$-dimensions. An action can be rendered {\it ghost-free} by requiring that no additional degrees of freedom other than the massless graviton are introduced. \footnote{More generally, one simply requires that the modified propagator is proportional to the propagator of GR. In such a case, it is possible, within certain limitations, to introduce an additional scalar propagating degree of a freedom and for the theory to remain ghost-free.} This is in spite of the fact that 
there are an infinite number of derivatives acting on the curvature. The action, in fact, remains generally covariant, giving rise to non-local graviton interactions, and has improved UV behaviour in
terms of higher loops~\cite{Talaganis:2014ida}, see also~\cite{Tomboulis,Modesto}.  

A further observation was  made by studying the {\it true} dynamical degrees
of freedom propagating between the spacetime region and the gravitational entropy of a self-gravitating system, containing a horizon. 
In Ref.~\cite{Conroy:2015wfa}, it was shown that the gravitational entropy of such a non-trivial, ghost-free gravitational action for a Schwarzschild's blackhole in 
$4$ dimensions is determined solely by the Einstein-Hilbert action. The higher-order corrections, at least up to quadratic in curvature,  do not 
contribute to the gravitational entropy.

The aim of this paper is two-fold. 

Firstly, we wish to:
\begin{itemize}
\item Extend the results of Refs.~\cite{Biswas:2011ar} to $D$-spacetime dimensions, by detailing the graviton propagator and the subsequent constraints for a {\it ghost-free} construction of gravity in $D$-dimensions around Minkowski background.
\item Generalise the entropy computation for a Schwarzschild-like blackhole to $D$-dimensions and show that the entropy for such a 
setup will be determinend solely by the Einstein-Hilbert action~\cite{Conroy:2015wfa}.
\end{itemize} 
And secondly, we intend to
\begin{itemize}
\item Study the gravitational entropy for a de-Sitter (dS) and Anti-de-Sitter (AdS) metric, again in $D$-spacetime-dimensions.
\item Discuss this result in the context of a non-singular bouncing cosmology.
\end{itemize}
Furthermore, note that in this paper we do not attempt to present the graviton propagator for (A)dS, leaving this for future study.

The paper is organised as follows: In section 2, we provide some technical details on how to obtain the graviton propagator in $D$-spacetime dimensions and study the 
ghost-free condition around Minkowski background. In Section 3, we discuss the general definition of gravitational entropy by following Wald's 
prescription. In Section 4, we present the gravitational entropy for a $D$ dimensional Schwarzschild-like blackhole. In Section 5, we study the gravitational
entropy for dS and AdS in $D$ dimensions, and in Section 6, we apply this result in a cosmological context.


\section{Infinite derivative gravity in $D$-dimensions}

The most general~\footnote{We are considering theories of gravity that are parity invariant and torsion-free.}, $D$-dimensional, non-local action of 
gravity that is quadratic in curvature can be expressed as a combination of the Einstein-Hilbert term and higher order terms~\footnote{The structure is 
very similar to Ref.~\cite{Biswas:2011ar}, which was derived in $4$ dimensions.}
\begin{eqnarray}\label{action1}
I^{tot} =\frac{1}{16\pi G_D}\int d^Dx \sqrt{-g}\left[ R +
\alpha \left(R {\cal F}_1(\Box)R\right.\right.
\nonumber \\
\left. \left. +R_{\mu\nu}{\cal F}_2(\Box)R^{\mu\nu} + R_{\mu\nu\lambda\sigma}{\cal
 F}_{3}(\Box)R^{\mu\nu\lambda\sigma}\right)
\right]\,,
\end{eqnarray}
where $G_D$ is the $D$-dimensional Newton's constant~\footnote{In $D$-dimension $G_D$ has dimension of $[G^{(D)}]=[G^{(4)}]L^{D-4}$ where L is unit length.}; 
$\alpha $ is a constant~\footnote{Note that for an arbitrary choice of $\mathcal{F}(\Box) $ at action level, $\alpha$ can be positive or negative as one can absorbs the sign into the coefficients $f_{i_{n}}$ contained within $\mathcal{F}(\Box)$ to keep the overall action unchanged, however $\alpha$ has to be strictly positive once we impose ghost-free condition.~\cite{Biswas:2005qr}   } with dimension of inverse mass squared; and $\mu,~\nu,~\lambda,~\sigma$
run from $0,~1,~2,\cdots  D-1$.  We briefly note that actions of higher order
in curvature are permitted by the general diffeomorphism, however, we restrict
ourself to actions of quadratic order, which capture all the quadratic  perturbations, around the Minkowski spacetime,
required for studying the graviton propagator~\cite{Biswas:2011ar}~\cite{Biswas:2013kla}. The form factors given by ${\cal F}_{i}(\Box)$ contain an infinite number of covariant derivatives, of the form:
\begin{equation}
{\cal F}_{i}(\Box)\equiv\sum_{n=0}^{\infty}f_{i_{n}}\biggl(\frac{\Box}{M^2}\biggr)^{n}\,,
\end{equation}
with constants $f_{i_{n}}$, and $\Box\equiv  g^{\mu\nu}\nabla_{\mu}\nabla_{\nu}$
is the D'Alembertian operator. The reader should note that, in our presentation, the function ${\cal F}_i(\Box)$ comes with an associated $D$-dimensional mass scale, 
$M \leq M_p = (1/\sqrt{(8\pi G_D)})$, which  determines the scale of non-locality in a quantum sense, see~\cite{Talaganis:2014ida}. 
The full classical equations of motion has been derived for action Eq.~(\ref{action1}) in $4$ dimensions in Ref.~\cite{Biswas:2013cha}, see also~\cite{Schmidt} for mathematical techniques.



In particular, let us consider perturbations around $D$-dimensional Minkowski spacetime  with metric tensor $\eta_{\mu\nu}$, such that $\eta_{\mu\nu}\eta^{\mu \nu}=D$, and where the perturbations are denoted by $h_{\mu\nu}$. One should also note that we are using {\it mostly plus} metric signature convention.

The $\mathcal{O}(h)$ expressions for the Riemann tensor, Ricci tensor and curvature scalar in $D$-dimensions are given by:
\begin{eqnarray}\label{ident-0}
R_{\mu\nu\lambda\sigma}=\frac{1}{2}(\partial_{[\lambda}\partial_{\nu}h_{\mu\sigma]}-\partial_{[\lambda}\partial_{\mu}h_{\nu\sigma]})\nonumber \\
R_{\mu\nu}=\frac{1}{2}(\partial_{\sigma}\partial_{(\nu}\partial^{\sigma}_{\mu)}-\partial_{\mu}\partial_{\nu}h-\Box
h_{\mu\nu})\nonumber \\
R=\partial_{\mu}\partial_{\nu}h^{\mu\nu}-\Box h .
\end{eqnarray}
The full action Eq.~(\ref{action1}) can then be expanded around Minkowski space, retaining terms up to
$\mathcal{O}(h^2)$:
\begin{eqnarray}\label{lin-act-0}
S_{q}= - \int d^D x \left[\frac{1}{2}h_{\mu \nu} \Box a(\Box) h^{\mu
\nu}+h_{\mu}^{\sigma} b(\Box) \partial_{\sigma} \partial_{\nu} h^{\mu \nu} \right.
\nonumber \\
\left.+h c(\Box)\partial_{\mu} \partial_{\nu}h^{\mu \nu} + \frac{1}{2}h \Box d(\Box)h \right.\nonumber \\
\left.+ h^{\lambda \sigma} \frac{f(\Box)}{2\Box}\partial_{\sigma}\partial_{\lambda}\partial_{\mu}\partial_{\nu}h^{\mu
\nu}\right] \,.
\end{eqnarray}
The above expression, along with the functional forms of $a(\Box)$, $b(\Box)$, $c(\Box)$, $d(\Box)$ and $f(\Box)$ 
remain the same as those of the $4$-dimensional case,  see Ref.~\cite{Biswas:2011ar,Biswas:2013kla}, and our Appendix A. One can easily note that 
\begin{equation}
f(\Box)=a(\Box)-c(\Box),
\end{equation}
and that the equation of motion satisfies the generalised Bianchi identities, for energy momentum tensor $\tau_{\mu\nu}$:
\begin{multline}\label{bianchi}
\nabla_{\mu}\tau^{\mu}_{\nu}=0=(c+d)\Box\partial_{\nu}h \\
 +(a+b)\Box h^{\mu}_{\nu,\mu}+(b+c+f)h^{\alpha\beta}_{,\alpha\beta\nu}\,.
\end{multline}
To find the graviton propagator in $D$-dimensions, we follow a similar projection operator prescription as found in Refs.~\cite{Van,Biswas:2011ar,Biswas:2013kla}, see also Appendix B. 

Thus the $D$-dimensional propagator is now given by
\begin{equation}
\label{fullprop}
\Pi(-k^{2})=\frac{\mathcal{P}^{2}}{k^{2}a(-k^{2})}+\frac{\mathcal{P}_{s}^{0}}{k^{2}\left(a(-k^{2})-(D-1)c(-k^{2})\right)}
 \end{equation} 
 Choosing $f(\Box)=0\Rightarrow
a(\Box)=c(\Box)$, so as not to introduce any scalar propagating degrees of freedom, we find 
\begin{equation}
\Pi = \frac{1}{k^2a(-k^2)} \left( \mathcal{P}^{2}-\frac{1}{D-2}\mathcal{P}_{s}^{0} \right).
\end{equation}
The form of $a(-k^2)$ should be such that it does not introduce any new propagating degree of freedom, and 
it was argued in Ref.~\cite{Biswas:2011ar,Biswas:2005qr} that the form of $a(\Box)$ should be an {\it entire function}, so as not to introduce
any pole in the complex plane, which would result in additional degrees of freedom in the momentum space~\footnote{Similar arguments for a propagator were also made in 
Refs.~\cite{Tomboulis} before ours.}.

Furthermore, the form of $a(-k^2)$ should be such that in the IR, for $k\rightarrow 0,~ a(-k^2) \rightarrow 1$, therefore recovering the propagator 
of GR in the $D$ dimensions. For $D=4$, the propagator has the familiar $1/2$ factor infront of the scalar part of the propagator. One such 
example of an {\it entire function} is~\cite{Biswas:2011ar,Biswas:2005qr}:
\begin{equation}\label{choice}
a(\Box) = e^{-\Box/M^2}\,,
\end{equation}
which has been found to ameliorate the UV aspects of gravity by removing the blackhole singularity within the linearised limit in a static 
configuration~\cite{Biswas:2011ar}, see also \cite{Tseytlin,Siegel:2003vt},
while recovering the Newtonian limit in the IR~\footnote{In principle one can take $a(\Box) = e^{(-\Box/M^2)^{n}}$, where $n$ is an integer, but a wrong choice of sign for instance $a(\Box)= e^{-(-\Box/M^2)^n}$ would not yield the correct Newtonian potential~\cite{Talaganis:2014ida}.}. In Refs.~\cite{Frolov-1,Frolov-2}, the authors have 
shown that the spherical collapse within a linearised limit does not form a curvature singularity in this class of models.

By imposing $f(\Box)=0$, i.e. $~a(\Box) = c(\Box)$, we obtain:
\begin{equation}\label{constraint}
2\mathcal{F}_{1}(\Box)+\mathcal{F}_{2}(\Box)+2\mathcal{F}_{3}(\Box)=0\,,
\end{equation}
which is independent of the number of spacetime dimensions. This result was first shown in Ref.~\cite{Biswas:2011ar} in $4$ dimensions.


\section{Gravitational entropy}

In the framework of Lagrangian field theory, Wald showed that one can find the gravitational entropy by varying the Lagrangian and subsequently finding the Noether current as
a function of an assigned vector field. By writing the corresponding Noether
charge, it has been shown that, for a static blackhole, the first law of thermodynamics can be satisfied and the entropy may be expressed by integrating the Noether
charge over a bifurcation surface of the horizon. In so doing, one must  choose the assigned vector field to be a horizon Killing vector, which has been normalised to unit surface gravity, see~\cite{Wald} and \cite{Kang}. 

In order to compute the gravitational entropy of the non-local theory outlined above, we impose a $D$-dimensional, static, 
homogenous and spherically symmetric metric, containing a {\it horizon}, of the form
\begin{equation}\label{metric1}
ds^2= -f(r) dt^2 + f(r)^{-1}dr^2+r^2d\Omega^{2}_{D-2}\,.
\end{equation}
This metric is equally applicable to (A)dS metric as it is to the static blackhole case, and results in the following entropy equation,
\begin{equation}\label{wald}
S_{W}=-8\pi\oint_{r=r_{H},t=const}\frac{\delta\mathcal{L}}{\delta R_{rtrt}}r^{D-2}d\Omega^{2}_{D-2}\,,
\end{equation}
where the bifurcation surface is at \(r=r_{H}\), $t=const$.  The area
of the horizon, \(A_{H}\), is given by integrating over the $(D-2)$-sphere as, 
\begin{equation}
A_{H}=\oint_{r=r_{H},t=const}r^{D-2}d\Omega^{2}_{D-2}.
\end{equation} 
In the following sections, we first derive the $D$-dimensional entropy, corresponding to  action Eq.~(\ref{action1}), before turning our attention to the case of a static blackhole and, subsequently, the dS and AdS cases.


By varying the Lagrangian density with respect to the Riemann tensor in \( (r,t)\) directions, as given in Eq.~(\ref{wald}), one may compute the entropy of the gravitational system described by the action Eq.~(\ref{action1}) and corresponding metric Eq.~(\ref{metric1}), as follows 
\begin{multline}\label{bh-entropy}
S_{W}=\frac{A_{H}}{4G_{D}}\bigl[1+\alpha(2 {\cal F}_1(\Box)R \\
 - {\cal F}_2(\Box)\times(g^{rr}R^{tt}+g^{rr}R^{rr})-4 {\cal F}_3(\Box)R^{rtrt})\bigr]
\end{multline}
It is convenient, for illustrative purposes, to decompose the entropy equation into its $(r,t)$ and spherical components.  For a static and axis-symmetric metric Eq.~(\ref{metric1}), we denote the \(r\) and \(t\) directions by the indices \(\{a,b\}\); and the spherical components by \(\{\bar m,\bar n\}\). 
As such, we express the curvature scalar as follows
 \begin{equation}
R=g^{\mu\nu}R_{\mu\nu}=g^{ab}R_{ab}+g^{\bar m\bar n}R_{\bar m\bar n},
\end{equation}
where $g_{ab}$ is a $2$-dimensional metric tensor accounting for the ${r,t}$ directions and $g_{\bar m\bar n}$ is a $(D-2)$-dimensional metric tensor, corresponding to the angular components, such that $$g^{\mu\nu}g_{\mu\nu}\equiv g^{ab}g_{ab}+g^{\bar m\bar n}g_{\bar m\bar n}=2+(D-2)= D.$$

Expanding the scalar curvature into Ricci and Riemann tensors, along with the properties of the static, spherically symmetric metric Eq.~(\ref{metric1}), allows us to express the relevant components of the entropy equation as follows
\begin{equation}
g^{rr}R^{tt}+g^{rr}R^{rr}=-g_{tt}R^{tt}-g_{rr}R^{rr}=-g^{ab}R_{ab}
\end{equation}
\begin{equation}
-4R_{rtrt}=2g^{ab}R_{ab}-2g^{ab}g^{\bar
m\bar n}R_{\bar ma\bar nb}
\end{equation}
Substitution into Eq.~(\ref{bh-entropy}), results in a decomposed $D$-dimensional entropy equation for the action~(\ref{action1}) in a static, spherically symmetric background
\begin{multline}\label{bh1}
S_{W}=\frac{A_{H}}{4G_{D}}[1+\alpha(2 {\cal F}_1(\Box)+ {\cal F}_2(\Box)+2 {\cal F}_3(\Box))\times\ \\
g^{ab}R_{ab}+2\alpha( {\cal F}_1(\Box)g^{\bar
m\bar n}R_{\bar
m\bar n}-
{\cal F}_3(\Box)g^{ab}g^{\bar
m\bar n}R_{\bar ma\bar nb})].
\end{multline}
This decomposed form is particular illustrative in the context of a static axis-symmetric blackhole.

  
\section{$D$-dimensional blackhole entropy }

For a spherically symmetric $D$-dimensional background, the angular components of the Ricci tensor are given by $R_{\theta_{n}\theta_{n}}=\sin^{-2}(\theta_{n}) R_{\theta_{n+1}\theta_{n+1}}$, where $\theta_n$ runs from $1$ to $D-2$, signifying each angular direction. Explicitly, for the given metric, $R_{\theta_{1}\theta_{1}}$ is given by
\begin{equation}
R_{\theta_{1}\theta_{1}}=(D-3)-(D-3)f(r)-r f^{\prime}(r)=0\,,
\end{equation}
for a vacuum solution, and solving the straightforward differential equation, one reveals
\begin{equation}
f(r)=1-\frac{\mu}{r^{D-3}}\,,
\end{equation}
where $\mu$ is a constant of integration. The form of this function and its associated metric, encompasses the $D$-dimensional analogue of the Schwarzschild solution, known as the Schwarzschild-Tangherlini metric  \cite{Tangherlini:1963bw}, with $\mu=\frac{16\pi G_{D}M}{(D-2)A_{D-2}}$,\cite{Emparan:2008eg}; and is an asymptotically $D$-dimensional Minkowski background.

Thus, when considering a Schwarzschild solution, all $R_{\theta_{i}\theta_{i}}$ components, will vanish. This is a consequence of the axisymmetric properties of the solution. Therefore, the entropy of a $D$-dimensional static and spherically symmetric metric with a horizon, yields:
\begin{equation}
S_{W}=\frac{A_{H}}{4G_{D}}[1+\alpha(2 {\cal F}_1(\Box)+ {\cal F}_2(\Box)+2 {\cal F}_3(\Box))\times
g^{ab}R_{ab}]\,.
\end{equation}
Combining the above with the constraint given in Eq.~\eqref{constraint}, results in the gravitational entropy of the modified sector vanishing entirely. Thus, the $D$-dimensional blackhole entropy corresponding to action Eq.~(\ref{action1})  is given solely by the Bekenstein-Hawking area law
\begin{equation}
S_{W}=\frac{A_{H}}{4G_{D}}\,.
\end{equation}
This simple observation ensures that,  in the context of a static, spherically symmetric metric, which asymptotes to Minkowski, 
the holographic nature of gravity is preserved in the IR. The higher-order corrections to the UV do not affect the gravitational entropy 
stored on the horizon, as long as the {\it only} propagating degrees of freedom are the massless graviton.  To some extent, one may be 
able to conjecture that the area-law of gravitational entropy is purely an IR concept of nature in such circumstances.

This is a powerful result. For instance, any arbitrary $f(R)$ theory of gravity would contribute towards the gravitational entropy for the above choice of metric. We can understand this very simply because such a class of theory would induce an extra scalar propagating degree of freedom other than 
the massless graviton, which would contribute towards the gravitational entropy.

In some sense, there is an intriguing connection between propagating degrees of freedom and the gravitational entropy at the horizon, at least in the context of a static,
spherically symmetric background, which asymptotes to Minkowski. This elegant result may not hold for a (A)dS background as we shall see below, since the form of the propagator 
for infinite derivative theory of gravity will be different for dS and AdS compared to Minkowski.


\section{\label{dS}$D$-Dimensional \((A)dS\) Entropy}

We now turn our attention to another class of metrics which contain an horizon, such as the (A)dS metrics, 
where the $D$-dimensional non-local action Eq.~\eqref{action1} must now be appended with a cosmological constant $\Lambda$ to ensure that (A)dS is a vacuum solution.  
\begin{eqnarray}
\label{action-main2}
I^{tot} =\frac{1}{16\pi G_{D}}\int d^Dx \sqrt{-g} \bigl[ R-2\Lambda\nonumber\\+\alpha \bigl(R\mathcal{F}_1(\Box)R+R_{\mu\nu}\mathcal{F}_2(\Box)R^{\mu\nu}
\nonumber \\
 + R_{\mu\nu\lambda\sigma}
\mathcal{F}_{3}(\Box)R^{\mu\nu\lambda\sigma}\bigr)
\bigr]\,.
\end{eqnarray}
The cosmological constant is then given by
\begin{equation}
\Lambda=\pm\frac{(D-1)(D-2)}{2l^{2}}\,,
\end{equation} 
where the positive sign corresponds to dS, negative to AdS, and hereafter, the topmost sign will refer to dS and the bottom to AdS. $l$ denotes the cosmological horizon.
The (A)dS metric can be obtained by taking 
 \begin{equation}
f(r)=\left(1\mp\frac{r^2}{l^2}\right)\,,
\end{equation} 
in Eq.~(\ref{metric1}).

Recalling the $D$-dimensional entropy Eq.~(\ref{bh1}),   we write, 
\begin{multline}\label{bh3}
S^{(A)dS}_{W}=\frac{A^{(A)dS}_{H}}{4G_{D}}[1+\alpha(2 {\cal F}_1(\Box)+ {\cal F}_2(\Box)+2
{\cal F}_3(\Box))\times\ \\
g^{ab}R_{ab}+2\alpha( {\cal F}_1(\Box)g^{\bar
m\bar n}R_{\bar
m\bar n}-
{\cal F}_3(\Box)g^{ab}g^{\bar
m\bar n}R_{\bar ma\bar nb})]\,,
\end{multline} 
where now \(A^{(A)dS}_{H}\equiv l^{D-2}A_{D-2}\), with \(A_{D-2}=(2\pi^{\frac{D-1}{2}})/\Gamma[\frac{D-1}{2}]\). Given the $D$-dimensional definitions of curvature in (A)dS background, 
\begin{multline}\label{curvatures}
R_{\mu\nu\lambda\sigma}=\pm\frac{1}{l^{2}}g_{[\mu\lambda}g_{\nu]\sigma},\quad
R_{\mu\nu}=\pm\frac{D-1}{l^{2}}g_{\mu\nu},\\ R=\pm\frac{D(D-1)}{l^{2}},
\end{multline} 
simple substitution reveals the gravitational entropy in (A)dS can be 
expressed as:
\begin{equation}\label{bh4}
S^{(A)dS}_{W}=\frac{A^{(A)dS}_{H}
}{4 G_{D}}(1\pm\frac{2\alpha  }{l^{2}}\{f_{1_0}D(D-1)+f_{2_0}(D-1)^{}+2f_{3_0}\}).
\end{equation} 

Note that $f_{i_{0}}$'s are now simply the leading constants of the functions ${\cal F}_i(\Box)$, due to the nature of curvature in (A)dS. 
In particular, in $4$-dimensions, the combination $12f_{1_0}+3f_{2_0}+2f_{3_0}$ is very different from that of the Minkowski space constraint, 
see Eq.~(\ref{constraint}), required for the massless nature of a graviton in any $D$ dimensions around Minkowski. Deriving the precise form of the
ghost-free constraint in (A)dS, is still an open problem for the above action Eq.~(\ref{action1}).


  
\section{Non-singular bouncing cosmology and holographic entropy} 

One of the applications of seeking (A)dS gravitational entropy for an infinite derivative theory of gravity is to understand 
the initial conditions for the Universe. It has been known for sometime that non-locality in gravity resolves the cosmological
singularity problem, at least in the context of homogeneous and isotropic metric, such as Friedmann-Robertson-Walker (FRW) background~\cite{Biswas:2005qr}.
Even small inhomogeneities, i.e. sub-~\cite{Biswas:2012bp} and super-Hubble~\cite{Biswas:2010zk} perturbations around such a non-singular bouncing solution are stable~\footnote{In order to resolve the cosmological singularity in an asymptotically Minkowski background, one would need
to depart from the condition $a(\Box) =c(\Box)$ in \eqref{fullprop}. However, as alluded to in footnote 1, the resulting propagator must be proportional to the GR propagator in order to be ghost-free. This is also reflected in an FRW background, 
where one would require an extra scalar degree of freedom besides the massless graviton~\cite{Biswas:2005qr}.}. 

A sub-class of the full action Eq.~(\ref{action-main2}) in $4$ dimensions is given by~\footnote{This action has also been proposed as a UV complete action for Starobinsky 
inflation~\cite{Craps,Chialva:2014rla}.}:
\begin{equation}
\label{actionred}
I_{R} =\frac{1}{16\pi G_4}\int d^4 x \sqrt{-g} \left( R-2\Lambda+\alpha R {\cal F}_1(\Box)R \right)\,,
\end{equation}
where $\alpha>0$ is required to ensure that gravity remains ghost-free.
A reduced action of this type has been studied in \cite{Conroy:2014dja}, where it was shown in a FRW spacetime 
(consequently dS), that null rays can be made past-complete without violating any relevant energy conditions~\cite{Conroy:2014dja}, thus replacing the cosmological 
singularity with a \emph{bounce} at $t=0$.



It was shown in~\cite{Biswas:2005qr} that a spacetime may be rendered ghost-free for the following choice of $\mathcal{F}(\Box)$:
\begin{equation}
{\cal F}_{1}(\Box)=\frac{e^{-\Box/M^{2}}-1}{\Box/M^2}.
\end{equation}
A well-defined background solution which would satisfy the 
equations of motion for the above action Eq.~(\ref{actionred}) is given by~\cite{Biswas:2010zk,Biswas:2012bp}
\begin{equation}
a(t)=a_0\cosh\left( \sqrt{\frac{\Lambda}{3}} t\right)\,,
\end{equation}
where typically  $\Lambda^{1/2} \approx M$, i.e. the scale of non-locality, $a_0$ is just a constant, and 
$H=l^{-1}=\sqrt{{\Lambda}/{3}}$. At the bounce, 
$\dot a=0,~\ddot a > 0$, where dot denotes time derivative with respect to physical time $t$.
The Universe loiters around a dS phase - the duration of which depends on how long the bouncing phase lasts around $t=0$.

It would indeed be worthwhile to ask what the gravitational entropy stored in a cosmological constant dominated Universe should be, at the time of the
bounce~\footnote{In the {\em classical } Einstein gravity this question is ill-defined - without violating the
energy conditions it is not possible to avoid the cosmological singularity.}. 
From Eq.~\eqref{bh4}, we read off the entropy equation for the action Eq.~\eqref{actionred} in de Sitter space \footnote{In AdS, the equivalent entropy of Eq. \eqref{dSentropyred}, is given by $S_{W}^{AdS}=\frac{A_{H}^{AdS}}{4G_{4}}\biggl(1+\frac{24\alpha}{l^{2}}\biggr)$.}
\begin{equation}
\label{dSentropyred}
S^{dS}_{R}=\frac{A^{dS}_{H}
}{4 G_{4}}\biggl(1-\frac{24\alpha  }{l^{2}}\biggr)\,.
\end{equation}

\section{Discussion and conclusion}

The first thing to note here is that modified gravitational entropy in a
cosmological constant dominated background is diminished with respect to
the entropy of Einstein's gravity. Furthermore, one notes that for a particular
value of the arbitrary dimensionful quantity $\alpha=\frac{l^2}{24}$, the
entropy vanishes entirely. This is in contrast to the blackhole case where
the \emph{UV modified} sector does not contribute to the gravitational entropy,
when no extra degrees of freedom are introduced into the system. In this
case, the possibility that the gravitational entropy will vanish in its entirety,
is allowed for. This is an intriguing outcome of infinite derivative non-local
gravity. 

It also raises the question: \emph{Could our Universe have begun its journey
with a zero gravitational entropy?} At the present moment, we merely speculate
on the notion of 
a zero gravitational entropy state at the \emph{bounce point} of cosmology.
A zero entropy state for any system would be equivalent to 
realising a \emph{ground state} of the system. In our case, it is the graviton
which realises its ground state in the presence of $\Lambda$ and non-local
gravity.
Could this lead to a new state of gravity such that our Universe would yield
a condensation of gravitons, at the moment of bounce, similar to the  
Bose-Einstein condensate with a zero entropy state~\cite{Rief}? 

Some of these issues are indeed  fundamental in nature and would perhaps
open up new vistas
towards understanding the nature of quantum aspects of gravity in extreme
conditions and indeed, the nature of thermodynamics around the bounce.

In summary, we first demonstrated that in requiring that an infinite derivative theory of gravity, up to quadratic in curvature, is covariant and ghost-free,  
a constraint on the form factors ${\cal F}_{i}$ is revealed, given by Eq.~(\ref{constraint}). The relationship holds in any arbitrary $D$ dimensions. This 
is an extension of previous works, see~\cite{Biswas:2011ar}, where the results were known to hold only in $4$ dimensions. We further obtained the graviton propagator 
for such infinite derivative theory of gravity in $D$ dimensions such that the graviton remains massless and free from tachyon and ghosts.

We then studied the gravitational entropy at the horizon for an infinite derivative theory of gravity for a $D$-dimensional blackhole solution, i.e. a static 
and axis-symmetric metric. We confirmed the area-law of gravitational entropy and reinstated the connection between the ghost-free condition for gravitons and 
the holographic nature of gravity in the IR. 

We computed the gravitational entropy for a $D$ dimensions A(dS) metrics for such an infinite derivative theory of gravity. Unlike the blackhole case, 
the gravitational entropy due to extra contributions from the UV does not vanish. As an application, we studied the gravitational entropy of a non-singular 
bouncing cosmology at the bounce point, where we speculated upon the vanishing of gravitational entropy in its entirety.

\section{Acknowledgement}
The authors would wish to thank Alex Koshelev and Tirthabir Biswas for numerous discussions.
AC is funded by STFC grant no ST/K50208X/1, AM is supported by the STFC grant ST/J000418/1 and ST is supported by a scholarship from the Onassis Foundation. 

\appendix
\section{\label{sec:A} Bianchi identities }

The $\mathcal{O}(h^2)$ part of the full action can be written as in Eq.~(\ref{lin-act-0})
\begin{multline}
S_{q}= - \int d^D x [\frac{1}{2}h_{\mu \nu} \Box a(\Box) h^{\mu
\nu}+h_{\mu}^{\sigma} b(\Box) \partial_{\sigma} \partial_{\nu} h^{\mu \nu}
\\
+h c(\Box)\partial_{\mu} \partial_{\nu}h^{\mu \nu} + \frac{1}{2}h \Box d(\Box)h+ h^{\lambda \sigma} \frac{f(\Box)}{2\Box}\partial_{\sigma}\partial_{\lambda}\partial_{\mu}\partial_{\nu}h^{\mu
\nu}].
\end{multline}
where we have, 
\begin{multline}
R \mathcal{F}_{1}(\Box)R=\mathcal{F}_{1}(\Box)[h\Box^2 h+h^{\lambda \sigma}\partial_{\sigma}\partial_{\lambda}\partial_{\mu}\partial_{\nu}h^{\mu
\nu} \\
-2h\Box \partial_{\mu}\partial_{\nu}h^{\mu \nu}] 
\end{multline}
\begin{multline}
R_{\mu \nu}\mathcal{F}_{2}(\Box)R^{\mu \nu}= \mathcal{F}_{2}(\Box)[\frac{1}{4}h
\Box^2 h+\frac{1}{4}h_{\mu \nu}\Box^2 h^{\mu \nu}\\
-\frac{1}{2}h_{\mu}^{\sigma}\Box \partial_{\sigma}\partial_{\nu}h^{\mu \nu}-\frac{1}{2}h\Box
\partial_{\mu}\partial_{\nu}h^{\mu \nu}+ \frac{1}{2}h^{\lambda \sigma}\partial_{\sigma}\partial_{\lambda}\partial_{\mu}\partial_{\nu}h^{\mu
\nu}]
\end{multline}
\begin{multline}
R_{\mu \nu \lambda \sigma} \mathcal{F}_{3}(\Box)R^{\mu \nu \lambda \sigma}=\mathcal{F}_{3}(\Box)[h_{\mu
\nu}\Box^2 h^{\mu \nu} \\
 -2h_{\mu}^{\sigma}\Box \partial_{\sigma}\partial_{\nu}h^{\mu
\nu}+h^{\lambda \sigma}\partial_{\sigma}\partial_{\lambda}\partial_{\mu}\partial_{\nu}h^{\mu
\nu}]
\end{multline}
it should be noted that no $\eta_{\mu \nu} \eta^{\mu \nu}$ contractions appear.
Hence, $a(\Box)$, $b(\Box)$, $c(\Box)$, $d(\Box)$ and $f(\Box)$
are as follows:
\begin{align}
a(\Box)&=1-\frac{1}{2}\mathcal{F}_{2}(\Box)\frac{\Box}{M^2}-2\mathcal{F}_{3}(\Box)\frac{\Box}{M^2}, \\
b(\Box)&=-1+\frac{1}{2}\mathcal{F}_{2}(\Box)\frac{\Box}{M^2}+2\mathcal{F}_{3}(\Box)\frac{\Box}{M^2},\\
c(\Box)&=1+2\mathcal{F}_{1}(\Box)\frac{\Box}{M^2}+\frac{1}{2}\mathcal{F}_{2}(\Box)\frac{\Box}{M^2},\\
d(\Box)&=-1-2\mathcal{F}_{1}(\Box)\frac{\Box}{M^2}-\frac{1}{2}\mathcal{F}_{2}(\Box)\frac{\Box}{M^2},\\
f(\Box)&=-2\mathcal{F}_{1}(\Box)\frac{\Box}{M^2}-\mathcal{F}_{2}(\Box)\frac{\Box}{M^2}-2\mathcal{F}_{3}(\Box)\frac{\Box}{M^2}.
\end{align}
We observe that 
\begin{align}
a(\Box)+b(\Box)&=0\,,
\\c(\Box)+d(\Box)&=0\,,
\\ b(\Box)+c(\Box)+f(\Box)&=0\,.
\end{align}
Assuming $f(\Box)=0$, we get that $a(\Box)=c(\Box)$ and hence the earlier
constraint, $2\mathcal{F}_{1}(\Box)+\mathcal{F}_{2}(\Box)+2\mathcal{F}_{3}(\Box)=0$.
The generalised Bianchi identities give
\begin{multline}\label{bianchi}
\nabla_{\mu}\tau^{\mu}_{\nu}=0=(c+d)\Box\partial_{\nu}h \\
 +(a+b)\Box h^{\mu}_{\nu,\mu}+(b+c+f)h^{\alpha\beta}_{,\alpha\beta\nu}\,,
\end{multline}
from which we can verify the constraints (A10-A12).

\section{\label{sec:B } Spin projection operators in $D$ dimensional Minkowski space}

 Now the spin projector
operators in $D$-dimension Minkowski space are, see~\cite{Van,Biswas:2013kla}
\begin{equation}
\mathcal{P}^{2}=\frac{1}{2}(\theta_{\mu \rho}\theta_{\nu \sigma}+\theta_{\mu
\sigma}\theta_{\nu \rho} ) - \frac{1}{D-1}\theta_{\mu \nu}\theta_{\rho
\sigma},
\end{equation}
\begin{equation}
\mathcal{P}^{1}=\frac{1}{2}( \theta_{\mu \rho}\omega_{\nu \sigma}+\theta_{\mu
\sigma}\omega_{\nu \rho}+\theta_{\nu \rho}\omega_{\mu \sigma}+\theta_{\nu
\sigma}\omega_{\mu \rho} ),
\end{equation}
\begin{equation}
\mathcal{P}_{s}^{0}=\frac{1}{D-1}\theta_{\mu \nu} \theta_{\rho \sigma},
\end{equation}
\begin{equation}
\mathcal{P}_{w}^{0}=\omega_{\mu \nu}\omega_{\rho \sigma},
\end{equation}
\begin{equation}
\mathcal{P}_{sw}^{0}=\frac{1}{\sqrt{D-1}}\theta_{\mu \nu}\omega_{\rho \sigma},
\end{equation}
\begin{equation}
\mathcal{P}_{ws}^{0}=\frac{1}{\sqrt{D-1}}\omega_{\mu \nu}\theta_{\rho \sigma},
\end{equation}
where
\begin{equation}
\theta_{\mu \nu}=\eta_{\mu \nu}-\frac{k_{\mu}k_{\nu}}{k^2}
\end{equation}
and
\begin{equation}
\omega_{\mu \nu}=\frac{k_{\mu}k_{\nu}}{k^2}.
\end{equation}
we have
\begin{equation}
a(\Box)h_{\mu \nu} \rightarrow\ a(-k^2)\left[\mathcal{P}^{2}+\mathcal{P}^{1}+\mathcal{P}_{s}^{0}+\mathcal{P}_{w}^{0}\right]h,
\end{equation}
\begin{equation}
b(\Box)\partial_{\sigma}\partial_{(\nu}h_{\mu)}^{\sigma} \rightarrow -b(-k^2)k^2\left[\mathcal{P}^1+2\mathcal{P}_{w}^{0}\right]h,
\end{equation}
\begin{multline}
c(\Box)(\eta_{\mu \nu}\partial_{\rho}\partial_{\sigma}h^{\rho \sigma}+\partial_{\mu}\partial_{\nu}h
 )\\ 
\rightarrow\  -c(-k^2)k^2 \left[2\mathcal{P}_{w}^{0}+\sqrt{D-1}\left(\mathcal{P}_{sw}^{0}+\mathcal{P}_{ws}^{0}\right)
\right]h,
\end{multline}
\begin{multline}
 \eta_{\mu \nu} d(\Box)h \\
\rightarrow\  d(-k^2)\left[(D-1)\mathcal{P}_{s}^{0}+\mathcal{P}_{w}^{0}+\sqrt{D-1}\left(\mathcal{P}_{sw}^{0}+\mathcal{P}_{ws}^{0}\right)
\right]h,
\end{multline}
\begin{equation}
f(\Box)\partial^{\sigma}\partial^{\rho}\partial_{\mu}\partial_{\nu}h_{\rho
\sigma} \rightarrow\ f(-k^2)k^4\mathcal{P}_{w}^{0}h.
\end{equation}
Hence,
\begin{equation}
ak^2\mathcal{P}^2h=\kappa \mathcal{P} ^2 \tau \Rightarrow\ \mathcal{P}^{2}h=\kappa\left(\frac{\mathcal{P}^2}{ak^2}\right)\tau,
\end{equation}
\begin{equation}
(a+b)k^2 \mathcal{P}^{1}h=\kappa\mathcal{P}^{1}\tau \Rightarrow\ \mathcal{P}^{1}\tau=0,
\end{equation}
\begin{equation}
(a+(D-1)d)k^2\mathcal{P}_{s}^{0}h+(c+d)k^2\sqrt{D-1}\mathcal{P}_{sw}^{0}h=\kappa\mathcal{P}_{s}^{0}\tau,
\end{equation}
\begin{equation}
(c+d)k^2\sqrt{D-1}\mathcal{P}_{ws}^{0}h+(a+2b+2c+d+f)k^2\mathcal{P}_{w}^{0}h=\kappa\mathcal{P}_{w}^{0}\tau.
\end{equation}
So, 
\begin{multline}
(a+(D-1)d)k^2\mathcal{P}_{s}^{0}h= \kappa \mathcal{P}_{s}^{0}\tau \\
\Rightarrow\ \mathcal{P}_{s}^{0}h=\kappa \frac{\mathcal{P}_{s}^{0}}{(a+(D-1)d)k^2}\tau,
\end{multline}
\begin{multline}
(a+2b+2c+d+f)k^{2}\mathcal{P}_{w}^{0}h= \kappa \mathcal{P}_{w}^{0} \tau\\
\Rightarrow\mathcal{P}_{w}^{0} h= \kappa \frac{\mathcal{P}_{w}^{0}}{(a+2b+2c+d+f)k^2}
\tau,
\end{multline}
where we have used the constraints given by (A10-A12), and note that the denominator corresponding to the $P_w^0$ spin projector vanishes so that there is no $w$-multiplet. Thus, the $D$-dimensional propagator is given by
\begin{equation}
\label{fullprop}
\Pi(-k^{2})=\frac{\mathcal{P}^{2}}{k^{2}a(-k^{2})}+\frac{\mathcal{P}_{s}^{0}}{k^{2}(a(-k^{2})-(D-1)c(-k^{2}))}\,.
 \end{equation} 
Assuming $f(\Box)=0\Rightarrow
a(\Box)=c(\Box)$, so as not to introduce any scalar propagating degrees of freedom, we find 
\begin{equation}
\Pi = \frac{1}{k^2a(-k^2)} ( \mathcal{P}^{2}-\frac{1}{D-2}\mathcal{P}_{s}^{0}).
\end{equation}

\end{document}